\documentclass[aps,prd,10pt,nofootinbib,twocolumn]{revtex4-1}
\usepackage{amsmath,amssymb,amsfonts,dsfont,mathrsfs,amsthm,mathtools}
\usepackage{graphicx}
\usepackage{hyperref}
\usepackage{pstricks}
\usepackage{epsfig}
\usepackage[english]{babel}
\hypersetup{linktocpage,colorlinks=true,urlcolor=blue,linkcolor=blue,citecolor=blue}

\begin{document}

\title{Spontaneous Lorentz violation and asymptotic flatness}

\author{Yuri Bonder}
\email{bonder@nucleares.unam.mx}
\author{Christian Peterson}

\affiliation{Instituto de Ciencias Nucleares, Universidad Nacional Aut\'onoma 
de M\'exico\\ Apartado Postal 70-543, Cd.~Mx., 04510, M\'exico}

\begin{abstract}
The Standard Model Extension (SME) is a generic parametrization for Lorentz violation and the phenomenological consequences of the minimal gravity sector of the SME are usually studied using a post-Newtonian expansion that requires spacetime to be asymptotically flat. However, there is a term in this sector for which these approximations are unable to make predictions; this is known as the $t$ puzzle. The present paper studies a model of spontaneous Lorentz violation in the minimal gravity sector of the SME in a static and spherically symmetric situation, when no additional matter fields are present. It is shown that, under the above mentioned assumptions, $t$ is the only term in the minimal gravity sector for which no asymptotically flat solutions exist. This stems from the fact that the $t$ term fixes the asymptotic behavior of all the pieces of the curvature tensor.
\end{abstract}

\maketitle

\section{Introduction}\label{Intro}

Modern physics is based on the principle of local Lorentz invariance, which states that identical experiments performed in different inertial frames yield identical results. As a consequence there are no preferred spacetime directions. Since local Lorentz invariance is one of the basic assumptions of modern physics, it should be tested empirically. In addition, within several quantum gravity proposals, the possibility that local Lorentz invariance could be violated has been suggested (see, e.g., Refs. \onlinecite{KosteleckySamuel,GambiniPullin}). Thus, searching for violations of local Lorentz invariance, or Lorentz violation, for short, could shed light on the quantum nature of gravity, which is one of the biggest mysteries in our field.

To search for Lorentz violation it is extremely useful to have a general parametrization, which is called the Standard Model Extension (SME) \cite{SME1,SME2,Kostelecky2004}. The SME is constructed as an effective field theory \cite{Rob}, therefore, its action contains that of general relativity and the Standard Model, plus all terms that can be built with the fields and the symmetries of these two theories, excluding, of course, local Lorentz invariance. Thus, any term in the Lorentz-violating part of the SME action is comprised by a Lorentz-violating operator that is contracted to an SME coefficient; the latter can be thought of as the parameters for Lorentz violation.

It should be mentioned that, to date, there is no empirical evidence of Lorentz violation. In contrast, there are experimental bounds on many SME coefficients \cite{DataTables}, some of which reach the Planck scale, which is generally regarded as the quantum-gravity scale. Another useful feature of the SME is that its action is naturally divided into sectors; this division is inherited from the structure of conventional physics. Examples of these sectors include the electromagnetic sector \cite{PhysRevD.80.015020}, the neutrino sector \cite{Neutrinos}, and the gravity sector \cite{Kostelecky2004,KOSTELECKY2016510,PhysRevD.103.024059}. In addition, there is a hierarchy on the SME action terms given by the operator units. For example, in flat spacetime and for explicit Lorentz violation (see below), one can only consider the part of the SME known as the minimal subsector that only contains operators, built with the standard-model fields, that have dimensions such that they are power-counting renormalizable. Notably, this definition cannot be directly extended to dynamical spacetimes. Yet, for concreteness, this paper studies what is known as the minimal part of the gravity sector of the SME, which is defined below through its action.
 
Local Lorentz invariance can be broken explicitly or spontaneously. The latter means that the action is invariant under local Lorentz transformations but the relevant solutions to the field equations are not. In other words, the SME coefficients are dynamical, but the potential favors solutions where these coefficients seem ``frozen.'' In the case of spontaneous Lorentz violation and when spacetime is dynamical, the metric equation of motion is automatically divergence free; this is not the case when Lorentz violation arises explicitly. Thus, explicit Lorentz violation restricts the SME coefficients to have a consistent metric equation of motion (see, e.g., Ref.~\onlinecite{Explicit,PhysRevD.101.064056}). This result motivated the SME community to assume that, when spacetime is dynamical, one should consider spontaneous Lorentz violation \cite{Kostelecky2004,PhysRevD.91.065034}. In addition, it has been suggested that the Nambu-Goldstone modes associated with spontaneous Lorentz violation could account for the presence of photons and gravitons \cite{KosteleckyBlumm1,KosteleckyBlumm2}.  

In this paper it is assumed that local Lorentz invariance is spontaneously violated. Then, the questions becomes: What should the action for the SME coefficients be? One possibility is to give conditions on the solutions and fix the dynamics perturbatively \cite{BaileyKostelecky2006}. With this method the post-Newtonian corrections to general relativity can be obtained, producing expressions that can be compared with experiments \cite{tpuzz2,tpuzz3,tpuzz4,PhysRevD.88.102001,PhysRevLett.112.111103,kostelecky2015constraints,PhysRevLett.119.201101,PhysRevD.97.024019,EscobarMartin} (for a review on these experiments see Ref.~\onlinecite{reviewbounds}). Crucially, in this method the solutions are restricted to be asymptotically flat since the asymptotic region is used to define the perturbative scheme. However, under these conditions, not all the coefficients produce physical effects; concretely, the $t$ term produces no effects, giving rise to the $t$ puzzle. Another alternative, which is adopted here, is to propose a ``natural'' action for the SME coefficients \cite{AltschulBaileyKostelecky2010,Gabriel}. This framework produces physically relevant solutions that are not required to be asymptotically flat, including those useful for cosmology \cite{Gabriel}, and it is suitable for studying all the coefficients, including $t$.

The main goal of the paper is to shed light on the role of asymptotic flatness in spontaneous Lorentz violation. Concretely, the paper studies the issue of whether the minimal gravity sector of the SME has nontrivial solutions that are static, spherically symmetric, and \emph{asymptotically flat}, when local Lorentz invariance is spontaneously violated. The symmetry assumptions are taken for simplicity and because these symmetries are usually assumed to study the gravitational environment outside stars and planets (including the Sun and the Earth), where most of the classical tests of gravity have been made.

The paper is organized as follows: in Sec.~\ref{Prel} the action and equations of motion are given. Also, the notation and conventions are specified in this section. Then, in Sec.~\ref{secT}, a detailed study of the $t$ term under the conditions that all fields are static and spherically symmetric is presented, and it is shown that there are no asymptotically flat solutions. Section \ref{secSyU} deals with other terms in the minimal gravitational SME sector assuming the same symmetries. It is argued that these cases do admit asymptotically flat solutions. Finally the interpretation of the results and other concluding remarks appear in Sec.~\ref{Conc}, and in the Appendices the Hamiltonian and some black hole solutions are studied for a different class of potentials.

\section{Preliminaries}\label{Prel}

The action for the minimal gravitational sector of the SME is presented in this section. Spacetime is assumed to be four-dimensional and the action of the theory is \cite{Kostelecky2004}
\begin{equation}\label{action}
S[g,k]=\int d^4 x \sqrt{-g}\left[ \frac{1}{16\pi}\left(R + k^{abcd}R_{abcd}\right)+\mathcal{L}_k(g,k)\right],
\end{equation}
where the notation and conventions of Ref.~\cite{Wald} are adopted. In particular, Latin indexes from the beginning of the alphabet are abstract indexes, implying that those expressions are valid in any coordinate system, and repeated indexes are contracted. The spacetime metric is denoted by $g_{ab}$, it has signature $(-,+,+,+)$, its determinant is denoted by $g$, and indexes are lowered and raised with the metric and its inverse, $g^{ab}$. In addition, the units are such that $c=1=G$. ${R_{abc}}^d$ is the Riemann tensor associated with a torsionless and metric-compatible derivative $\nabla$, $R_{ab}\equiv{R_{acb}}^c$ is the Ricci tensor, $R\equiv{R_a}^a$ is the curvature scalar and ${W_{abc}}^d$ denotes the Weyl tensor, which is the completely traceless part of ${R_{abc}}^d$. Moreover, $k^{abcd}$ is the SME coefficient for the SME subsector under consideration; this coefficient has all the index symmetries of the Riemann tensor, thus having $20$ independent components. Some clarifications regarding the terminology used throughout the paper are also necessary: vacuum means that there are no additional fields besides $g_{ab}$ and $k^{abcd}$ and terms like the second term in the action \eqref{action} are called nonminimal coupling terms between the coefficients and the metric. Finally, $n$ indexes in between parenthesis (brackets) denotes the completely symmetric (antisymmetric) part with a $1/n!$ factor.

The Lagrangian for the SME coefficient, $\mathcal{L}_k(g,k)$, is expected to have a kinetic term and a potential such that $k^{abcd}$ produces spontaneous Lorentz violation. For concreteness, a particular form of such Lagrangian is chosen as
\begin{equation}\label{Lagk}
\mathcal{L}_k(g,k)=-\frac{1}{2}(\nabla k)^2 - V(k^2),
\end{equation}
where $(\nabla k)^2\equiv \nabla_e k_{abcd}\nabla^e k^{abcd}$ and $k^2\equiv k_{abcd} k^{abcd}$. Notice that the signs of these terms are set in such a way that the theory, in a flat spacetime, produces a Hamiltonian that is bounded from below for static solutions (see Appendix \ref{Hamilto}). In addition, the potential should be responsible of generating spontaneous Lorentz violation, thus, it could be taken as a generalization of the well-known Mexican hat potential
\begin{equation}\label{potential}
    V=a(k^2-b)^2,
\end{equation}
where $a>0$ and $b\neq 0$ are parameters; the fact that $b\neq 0$ produces spontaneous Lorentz violation since the solutions at the bottom of the potential correspond to $k^2=b$, and a nonzero $k^{abcd}$ typically picks preferred spacetime directions. Note that the theory described by the Lagrangian \eqref{Lagk} can be considered as a generalization of the Bumblebee models  \cite{PhysRevD.77.125007,PhysRevD.79.029902,BonderEscobar} where spontaneous Lorentz violation is produced by a vector field and which could generate other interesting phenomena (see Ref. \cite{PhysRevD.99.024042}).

It is useful to irreducibly decompose $k^{abcd}$ in such a way that \cite{Kostelecky2004}
\begin{equation}\label{irred1}
 k^{abcd}R_{abcd}=-u R + s^{ab}R^{\rm T}_{ab} + t^{abcd}W_{abcd}
\end{equation}
with $R^{\rm T}_{ab}$ denoting the traceless part of the Ricci tensor. The sign in the first term on the right-hand side is clearly conventional. It is easy to verify that $s^{ab}$ is symmetric and traceless, thus having nine independent components. Also, $t^{abcd}$, which is the completely traceless part of $k^{abcd}$, has all the index symmetries of $k^{abcd}$ and ten independent components. The last independent component of $k^{abcd}$ is the double trace $u$. It is possible to verify that, for Eq.~\eqref{irred1} to hold, it is necessary to make the following identifications:
\begin{eqnarray}
u&=&-\frac{1}{6}{k_{ab}}^{ab},\\
s^{ab}&=&2{k_c}^{acb}-\frac{1}{2} g^{ab} {k_{cd}}^{cd},\\
t^{abcd}&=&k^{abcd}-\left(g^{a[c} {k_e}^{d]eb} -g^{b[c} {k_e}^{d]ea}\right)\nonumber\\
&&+\frac{1}{3}g^{a[c}g^{d]b} {k_{ef}}^{ef}.
\end{eqnarray}
 
\section{The $t$ coefficient}\label{secT}

This section concerns the $k^{abcd}=t^{abcd}$ case, \textit{i.e.}, it is assumed that $u=0=s^{ab}$. Importantly, these restrictions on the SME coefficients are set before the action variation is calculated. Recall that this is the case that produces no effects in the post-Newtonian limit \cite{BaileyKostelecky2006}. The first task is to find the equations of motion, which is done next.

\subsection{Equations of motion}

To obtain the equations of motion, the stationary variation of the action \eqref{action}, with $k^{abcd}=t^{abcd}$, must be calculated. Attention is first set on the metric variation. Clearly, the first term in the action produces an Einstein tensor (plus boundary terms that are neglected throughout the manuscript; for a discussion on such terms see Ref.~\onlinecite{Bonder2015}). The second term has three parts: the variation of $\sqrt{-g}$, the variation of the metric that lowers the index in ${R_{abc}}^d$, and the variation of the Riemann tensor. The first piece produces $-\sqrt{-g}t^{cdef}R_{cdef} g_{ab}/2$, and in this expression the Riemann tensor can be replaced by the Weyl tensor using the trace free property of $t^{abcd}$. The second piece yields $-\sqrt{-g}R_{cde(a}{t^{cde}}_{b)}$, which, because its index structure, has contributions of the Weyl and Ricci tensors. The third part is obtained by using $\delta {R_{abc}}^d=-g^{de}\nabla_{[a}(\nabla_{b]} \delta g_{ec}+\nabla_{|c|} \delta g_{b]e}-\nabla_{|e|} \delta g_{b]c})$, which implies
\begin{equation}
\sqrt{-g}{k^{abc}}_d\delta {R_{abc}}^d=-2\sqrt{-g} (\nabla_{c} \nabla_{d} {{t^c}_{(ab)}}^d )\delta g^{ab},
\end{equation}
where $\delta g^{ab}=-g^{ac}g^{bd}\delta g_{cd}$ is used. Note that, in the last equation, the derivatives act symmetrically. 

After collecting all the terms, it can be verified that the metric equation of motion is
\begin{equation}
G_{ab}-\frac{g_{ab}}{2}t^{cdef}W_{cdef} - R_{cde(a}{t^{cde}}_{b)} - 2 \nabla_{c} \nabla_{d} {{t^c}_{(ab)}}^d=8\pi T_{ab}, \label{modEinstein}
\end{equation}
where
\begin{eqnarray}
T_{ab}&\equiv&-\frac{2}{\sqrt{-g}} \frac{\delta (\sqrt{-g}\mathcal{L}_k)}{\delta g^{ab} }\nonumber\\
&=&g_{ab}\left[-\frac{1}{2}(\nabla t)^2-a(t^2-b)^2\right] + \nabla_a t^{cdef} \nabla_b t_{cdef}\nonumber\\
&&+4\nabla_c\left[ {t_{def}}^c \nabla_{(a}{t^{def}}_{b)}- t_{def(a}\nabla_{b)}t^{defc}\right]\nonumber\\
&&+4 t_{def(a}\nabla_c\nabla^c {t^{def}}_{b)}-16a(t^2-b) t_{cde(a}{t^{cde}}_{b)}.\nonumber\\
&& \label{ExpTab}
\end{eqnarray}
To calculate $T_{ab}$ it is necessary to consider the contribution of $\sqrt{-g}$, and the fact that there are several metrics in $t^2\equiv t_{abcd}t^{abcd}$ and in $(\nabla t)^2\equiv\nabla_e t_{abcd}\nabla^e t^{abcd}$. Also, it is necessary to consider the presence of the metric-compatible connection ${\Gamma^c}_{ab}$ that appears when the derivative operator acts on $t^{abcd}$, which satisfies $\delta{\Gamma^c}_{ab}=g^{cd}(\nabla_a \delta g_{db}+\nabla_b \delta g_{ad}-\nabla_d \delta g_{ab})/2$.

The variation with respect to $t^{abcd}$ can be easily obtained: the second term in the action \eqref{action} produces $W_{abcd}$, the kinetic term in $\mathcal{L}_k$ must be integrated by parts to yield a d'Alembertian acting on $t_{abcd}$, and the variation of the potential $V=a(t^2-b)^2$ is straightforward. Hence, the equation of motion for $t^{abcd}$ is
\begin{equation}\label{EOMt}
W_{abcd}=16\pi \left[-\nabla_e \nabla^e t_{abcd} + 4 a(t^2-b) t_{abcd}\right].
\end{equation}

Given that the action variation is involved, it is useful to check the result. This can be done by calculating $\nabla_a T^{ab}$ using three different methods. The first and most obvious method is to directly calculate the divergence of Eq.~\eqref{ExpTab}. Using Eq.~\eqref{EOMt}, this divergence can be cast into the form
\begin{equation}\label{nablaT}
8\pi \nabla^aT_{ab} =-\frac{1}{2}W_{cdef} \nabla_b t^{cdef}-2\nabla_a\left(R_{cdeb} t^{cdea} \right).
\end{equation}
The second method is to calculate the divergence of the modified Einstein equation, Eq.~\eqref{modEinstein}, using the fact that Einstein tensor is divergence-free. These computations are long and involve finding pairs of derivatives acting antisymmetrically, which can be replaced by the Riemann tensor. For example, the divergence of the third term on the right-hand side of Eq.~\eqref{modEinstein} yields
\begin{equation}
2\nabla^a\nabla_c \nabla_d{{t^c}_{(ab)}}^d = R_{cdeb} \nabla_a t^{cdea}- \nabla^a \left( R_{cde[a} {t^{cde}}_{b]}\right).
\end{equation}
Again, it can be shown that the divergence of Eq.~\eqref{modEinstein} produces Eq.~\eqref{nablaT}. The third method uses the fact that $ \nabla^aT_{ab}$ can be independently calculated by requiring the last term of the action \eqref{action} to be invariant under diffeomorphisms, which must be the case because all the fields are dynamical \cite{Cristobal1,Cristobal2}. Of course, this last method also leads to Eq.~\eqref{nablaT}, suggesting that the equations of motion \eqref{modEinstein}-\eqref{EOMt} are consistent.

\subsection{Symmetries}

Typically, finding generic solutions of gravity theories is extremely hard. Thus, people consider symmetric situations where the equations simplify. This is also the case in gravity theories with Lorentz violation where symmetry assumptions have allowed some groups to find interesting solutions \cite{PhysRevD.101.064056,Barausse}. In this paper we take a similar approach. In this section the conditions that all fields are static and spherically symmetric are implemented.

Spherical symmetry is defined by the existence of three spacelike Killing vector fields that satisfy the $\mathfrak{so}(3)$ algebra. On the other hand, a static solutions means that there exists a timelike Killing vector field that is hypersurface orthogonal \cite{Wald}. The Schwarzschild coordinates $(t,r,\theta,\varphi)$ are extremely useful in these conditions as they are adapted to these symmetries and are regular in the spacetime region under consideration (besides from the well-known problems with the angular coordinates). Moreover, in these coordinates the hypersurface orthogonality condition can be thought of as an invariance under $t \to-t$. The most general metric that satisfies these symmetries can be written as
\begin{equation}\label{Metrica}
 \text{d}s^2 = -f\text{d}t^2 + h\text{d}r^2 + r^2(\text{d}\theta^2 +\sin^2\theta \text{d} \varphi^2),
\end{equation}
where $f=f(r)$ and $h=h(r)$ are positive functions of the radial coordinate (note that $r>0$).

For $t^{abcd}$ to be compatible with these symmetries it is necessary that its Lie derivatives along all the above-discussed Killing vector fields vanish \cite{Frolov}. It can be verified that, under these symmetries, $t^{abcd}$ has only one degree of freedom \cite{PhysRevD.101.064056}. Concretely, the nonvanishing components of $t^{abcd}$ are
\begin{eqnarray*} 
 t^{trtr} &=& T, \\
 t^{t \theta t \theta} & = &\sin^2\theta t^{t \varphi t \varphi}= \frac{-h T}{2r^2}, \\
 t^{r \theta r \theta} &=& \sin^2\theta t^{r \varphi r \varphi}= \frac{f T}{2r^2}, \\
 t^{\theta\varphi\theta\varphi} &=& -\csc^2\theta \frac{f h T}{r^4}, 
 \end{eqnarray*}
where $T=T(r)$ is an arbitrary function. Notably, it is possible to show that $t^2=12 f^2 h^2 T^2$, which is non-negative. Thus, by inspecting the form of the potential, one can conclude that $b>0$.

Importantly, under the assumed symmetries, there are three independent equations of motion for the three independent functions $f$, $h$, and $T$. One could argue that there are four independent equations of motion: one from the $t^{abcd}$ equation \eqref{EOMt} and three equations from Eq. \eqref{modEinstein}, namely, the $tt$, $rr$, and $\theta\theta$ components (the nondiagonal components vanish due to the symmetries and the $\phi\phi$ equation is proportional to the $\theta\theta$). However, the divergence of the modified Einstein equation only holds when Eq. \eqref{EOMt} is valid. Thus, Eq. \eqref{EOMt} can be considered as a consequence of Eq. \eqref{modEinstein}. In the next subsection, this system of equations is solved in the asymptotic region.

\subsection{Asymptotic flatness}

Asymptotically flat spacetimes describe the geometry of isolated gravitational sources and they provide us with formal definitions that are extremely useful, like that of the total spacetime energy (see, e.g., \onlinecite[chapter 11]{Wald}). The formal and covariant definition of an asymptotically flat spacetime \cite{ashtekarhansen,ashtekar1980} is highly nontrivial and it involves compactifying spacetime through a conformal transformation and immersing it into a larger pseudo-Riemannian manifold. This procedure allows one to add a boundary to the original spacetime, which can be divided into several pieces. Perhaps the most useful parts are the future and past null infinities, which represent the ``place'' that light rays will eventually reach, or where light rays were emitted infinitely into the past. However, here we focus on another piece of such a boundary that is known as spatial infinity, $i^0$, and which, loosely speaking, is the region that is spatially related to the source and infinitely far from it. 

Crucially, in the situation at hand there are preferred coordinates $(t,r,\theta,\phi)$ that are adapted to the symmetries. Thus, taking the limit as one approaches $i^0$ amounts to taking the $r\to \infty$ limit. What is more, due to the symmetries under consideration, this limit is direction independent.

The most conservative assumption for the metric to be asymptotically flat is that
\begin{eqnarray}
f=1+\frac{f_1}{r}+\frac{f_2}{r^2}+O\left(\frac{1}{r^3}\right), \label{exp1}\\
h=1+\frac{h_1}{r}+\frac{h_2}{r^2}+O\left(\frac{1}{r^3}\right), \label{exp2}
\end{eqnarray}
with $f_1$, $f_2$, $h_1$, and $h_2$ constants. Something similar must be assumed for $T$, however, in this case the constant term may take a value different from $1$. This is because the constant term in the potential effectively acts as a cosmological constant and theories with such a constant do not contain asymptotically flat solutions, but solutions that are asymptotically (anti-)de Sitter \cite{Ashtekar_2000,Dolan_2019}. Thus, the constant part of the potential must be cancelled by $t^{abcd}$. Taking this into the account one can assume
\begin{equation}
T=T_0+\frac{T_1}{r}+\frac{T_2}{r^2}+O\left(\frac{1}{r^3}\right), \label{exp3}
\end{equation}
where $T_0$, $T_1$, and $T_2$ are constants.

The expansions \eqref{exp1}-\eqref{exp3} can now be used in the independent equations of motion to look for solutions. Concretely, the $tt$ and $rr$ components of the modified Einstein equation \eqref{modEinstein} and the $trtr$ component of the $t^{abcd}$ equation \eqref{EOMt} are used. When this substitutions are done, the equations of motion become series in $1/r$, which can be solved recursively; notably, these are algebraic equations.

The simplest path to solve all the equations is to start with the series arising form Eq.~\eqref{EOMt}, which becomes
\begin{eqnarray}
0&=&8 a   T_0 \left(b-12 T_0^2\right)\nonumber\\
&&+\frac{8 a }{r}  \left[2 T_0 \left(b-24 T_0^2\right)
   (f_1+h_1)+T_1 \left(b-36 T_0^2\right)\right]\nonumber\\
&&-\frac{4 }{r^2}  \Big\{48   a T_0^3 \left[3 f_1^2+8 f_1 h_1+2 (f_2+h_2)+3 h_1^2\right]\nonumber\\
   &&+72 a T_0^2 [4 T_1   (f_1+h_1)+T_2]\nonumber\\
   &&-2 abT_0 \left(f_1^2+4
   f_1 h_1+2 (f_2+h_2)+h_1^2\right)\nonumber\\
   &&+72 a T_0T_1^2+3T_0\nonumber\\
   &&-2 ab \left[2 T_1 \left(f_1+h_1\right)+T_2\right]
   \Big\}+O\left(\frac{1}{r^3}\right).
\end{eqnarray}
From the first term one gets $T_0$, which is then used in the second term to get $T_1$, and so on. It can also be seen from the first term in this expansion that there are several solutions, in this case $T_0=0$ and $T_0=\pm \sqrt{b/12}$. The first solution implies $T_1=0=T_2$, or equivalently $T=O(1/r^3)$. The other solutions yield
\begin{eqnarray}
T_1&=&\mp \frac{\sqrt{b} (f_1+h_1)}{2 \sqrt{3}},\\
T_2&=&\pm \frac{4 ab \left(f_1^2+f_1h_1-f_2+h_1^2-h_2\right)-3}{8a \sqrt{3b}}.
\end{eqnarray}

At this stage it is possible to use the components of the modified Einstein equation \eqref{modEinstein} to solve for the metric functions. In the case where $T=O(1/r^3)$, the $tt$ component of Eq.~\eqref{modEinstein} becomes
\begin{equation}
    0=-a b^2 -\frac{a b^2 f_1 }{r}-\frac{a b^2 f_2}{r^2}+O\left(\frac{1}{r^3}\right),
\end{equation}
which has no solution under the working hypothesis (recall that $a>0$, $b>0$). Similar expressions are obtained if other component of Eq.~\eqref{modEinstein} are used. This implies that $T=O(1/r^3)$ is not a solution of all the independent equations. When the other solutions are inserted into the $tt$ component of Eq.~\eqref{modEinstein} one gets
\begin{equation}
   0= \frac{24 \pi b \pm \sqrt{3b} }{r^2}+O\left(\frac{1}{r^3}\right),
\end{equation}
which, again, has no solution unless $b=0$. Thus, it is possible to conclude that there are no static and spherically-symmetric solutions of the theory that are asymptotically flat. An explanation of this result is provided after the $s^{ab}$ and $u$ coefficients are analyzed, which is done in the next section.

\section{Other coefficients}\label{secSyU}

In this section the $u$ and $s^{ab}$ coefficients are analyzed under the assumptions used to study $t^{abcd}$ in Sec. ~\ref{secT}, namely, under the conditions that the solutions are static, spherically symmetric, and asymptotically flat. It should be noted that the $k^{abcd}$ decomposition into $t^{abcd}$, $s^{ab}$ and $u$ depends on $g_{ab}$. In the case studied above when $s^{ab}=0=u$, all the terms in $k^{abcd}$ that depend on the metric vanish. However, in the cases that are studied in this section there is an explicit metric dependency that has to be considered when the action  variations are calculated. Therefore, a new action for each of the remaining coefficients is proposed; these actions have the same form than the action \eqref{action}.

\subsection{The \textit{u} coefficient}

The action in this case is
\begin{equation}\label{action_u}
S[g,u]=\int d^4 x \sqrt{-g}\left[ \frac{1}{16\pi}\left(R - uR\right)+\mathcal{L}_u(g,u)\right],
\end{equation}
where
\begin{equation}
\mathcal{L}_u(g,u)=-\frac{1}{2}(\nabla u)^2 - a(u^2-b)^2.
\end{equation}
Again, $(\nabla u)^2\equiv \nabla_a u\nabla^a u$, and $a>0$, $b\neq 0$ are theory's parameters.

The equations of motion are
\begin{eqnarray}
G_{ab}(1+u)-\nabla_a\nabla_bu +g_{ab}\nabla_c\nabla^c u=8\pi T_{ab}, \label{modEinstein_u}\\
R=16\pi \left[-\nabla_c\nabla^c u+4a(u^2-b)u\right],\label{EOM_u}
\end{eqnarray}
where
\begin{equation}
T_{ab}=\nabla_a u \nabla_b u-\frac{1}{2}g_{ab} (\nabla u)^2- a g_{ab} (u^2-b)^2.\label{Tab_u}
\end{equation}
These equations are checked by calculating the divergence of $T_{ab}$ with the above-described independent methods: calculating directly from Eq.~\eqref{Tab_u}, taking the divergence of Eq.~\eqref{modEinstein_u}, and using invariance under diffeomorphism of the last term in the action \eqref{action_u}.

Under the assumption that all the fields are static and spherically symmetric, $u$ becomes a function of $r$, namely, $u=U(r)$. Since $u^2\geq0$, then $b>0$. In the $r\to \infty$ region one can assume that
\begin{eqnarray}
U=U_{0}+\frac{U_{1}}{r}+\frac{U_{2}}{r^2}+O\left(\frac{1}{r^3}\right).
\end{eqnarray}
where $U_0$, $U_1$, and $U_2$ are constants. When inserting these expressions and the metric expansions \eqref{exp1}-\eqref{exp2} into the equations of motion \eqref{modEinstein_u}-\eqref{EOM_u} one obtains equations that can be solved order by order in $1/r$. It is convenient to start by solving Eq.~\eqref{EOM_u}. To zeroth order, this equation is
\begin{equation}
    0= U_0 (b - U_0^2),
\end{equation}
which has three solutions. After solving to all the relevant orders in $1/r$ it is possible to argue that the solutions are the trivial solution $U=O\left(1/r^3\right)$ and $U=\pm\sqrt{b}+O\left(1/r^3\right)$. The last solutions correspond to $u$ lying at the bottom of the potential.

These solutions must now be inserted into Eq.~\eqref{modEinstein_u}. The trivial solution, when inserted into the $tt$ component of Eq.~\eqref{modEinstein_u}, produces an equation that has no solution under the working hypothesis. On the other hand, the solutions at the bottom of the potential automatically solve all the components of Eq.~\eqref{modEinstein_u}. Therefore, the solutions at the bottom of the potential produce solutions of all the equations of motion that are static, spherically symmetric, and asymptotically flat.

\subsection{The \textit{s} coefficient}

The action for this coefficient is
\begin{equation}\label{action_s}
S[g,s]=\int d^4 x \sqrt{-g}\left[ \frac{1}{16\pi}\left(R + s^{ab}R^{\rm T}_{ab}\right)+\mathcal{L}_s(g,s)\right],
\end{equation}
where 
\begin{equation}
\mathcal{L}_s(g,s)=-\frac{1}{2}(\nabla s)^2 - a(s^2-b)^2.
\end{equation}
Once again, $(\nabla s)^2\equiv \nabla_c s_{ab}\nabla^c s^{ab}$,  $s^2\equiv s_{ab} s^{ab}$, and $a>0$, $b\neq 0$ are parameters.

The equations of motion take the form:
\begin{eqnarray}
G_{ab}&-&\frac{g_{ab}}{2}s^{cd}R_{cd}  - 2 \nabla_{c} \nabla_{(a} {s_{b)}}^c+\frac{1}{2}g_{ab}\nabla_c \nabla_d s^{cd}\nonumber\\
&&+\frac{1}{2} g^{cd}\nabla_c \nabla_d s_{ab}=8\pi T_{ab}, \label{modEinstein_s}\\
R^{\rm T}_{ab}&=&16\pi \left[-g^{cd}\nabla_c\nabla_d s_{ab}+4a(s^2-b)s_{ab}\right],\label{EOM_s}
\end{eqnarray}
where
\begin{eqnarray}
T_{ab}
&=&g_{ab}\left[-\frac{1}{2}(\nabla s)^2-a(s^2-b)^2\right] +8a(s^2-b) s_{c(a}{s^{c}}_{b)}\nonumber\\
&& +2\nabla_c\left[s^{cd}\nabla_{(a} s_{b)d} - s_{d(a}\nabla_{b)}s^{cd}\right]\nonumber\\
 &&+2 g^{cd}{s^e}_{(a}\nabla_{|c} \nabla_{d|} s_{b)e},\label{Tab_s}
\end{eqnarray}
which are checked by obtaining $\nabla^a T_{ab}$ with the above-described methods.

Implementing the assumption that the fields are static and spherically symmetric implies that the only nonzero components of $s^{ab}$ are $s^{tt}$, $
s^{rr}$, and $s^{\theta\theta}=\sin^2 \theta s^{\phi\phi}$. In addition, the traceless property of $s^{ab}$ allows one to fix one of the remaining components in terms of the other. Thus, the nonzero components can be chosen as
\begin{eqnarray}
s^{tt}&=&S_1,\\
s^{rr}&=&S_2,\\
s^{\theta\theta}&=&\sin^2 \theta s^{\phi\phi}= \frac{f S_1 - h S_2}{2 r^2},
\end{eqnarray}
where $S_1$ and $S_2$ are arbitrary functions of $r$. Notice that $s^{ab}$ has two independent functions. Moreover, it is possible to show that 
\begin{equation}
    s^2=f^2 S_1^2 + h^2 S_2^2 + \frac{1}{2}(f S_1- h S_2)^2
\end{equation}
which is non-negative. This implies that $b>0$.

Asymptotically, the $s^{ab}$ functions can be assumed to be
\begin{eqnarray}
S_1=S_{1,0}+\frac{S_{1,1}}{r}+\frac{S_{1,2}}{r^2}+O\left(\frac{1}{r^3}\right),\\
S_2=S_{2,0}+\frac{S_{2,1}}{r}+\frac{S_{2,2}}{r^2}+O\left(\frac{1}{r^3}\right),
\end{eqnarray}
where $S_{i,j}$, with $i=1,2$, $j=0,1,2$, are constants. When these expansions and Eqs.~\eqref{exp1}-\eqref{exp2} are inserted into the equations of motion one obtains equations that can be solved order by order in $1/r$. It is convenient to begin by solving the $s^{ab}$ equation of motion, Eq.~\eqref{EOM_s}, which has two components. The $tt$ component can be used to find $S_{1,j}$. The zeroth order equation is
\begin{eqnarray}
0&=&S_{1,0} \left(2 b-3 S_{1,0}^2+2 S_{1,0} S_{2,0}-3 S_{2,0}^2\right),
\end{eqnarray}
which is cubic in $S_{1,0}$, thus, there are three different solutions to $S_{1,j}$. One of the solutions is the trivial solution $S_1=O\left(1/r^3\right)$ and the other two solutions differ by some signs.

These solutions are used in the $rr$ component of Eq.~\eqref{EOM_s} to find $S_{2,j}$. This procedure produces five solutions. The fact that there are no six independent equations, as one could expect, stems from a degeneracy in the trivial solution $S_1=O\left(1/r^3\right)=S_2$. The other four solutions are
\begin{eqnarray}
S_{1,0}&=& 0,\label{sols11}\\
S_{1,1}&=& 3 S_{2,1}\pm\sqrt{6} \sqrt{b}h_1,\\
S_{1,2}&=&3 \left[(9h_1-f_1) S_{2,1}+S_{2,2}\right]\pm\left(\frac{3}{2}\right)^{3/2}  \frac{\left(1+8 a S_{2,1}^2\right)}{a \sqrt{b}}\nonumber\\
   &&\pm \sqrt{6b} \left[4 h_1^2-f_1h_1+h_2\right],\\
S_{2,0}&=& \pm\sqrt{\frac{2b}{3}},\label{sols12}
\end{eqnarray}
and 
\begin{eqnarray}
S_{1,0}&=& \pm\frac{1}{2} \sqrt{3b} ,\label{sols21}\\
S_{1,1}&=&\pm \frac{f_1}{2}\sqrt{3b} ,\\
S_{1,2}&=&-\frac{3h_1 S_{2,1}}{8}\pm \frac{\sqrt{3b}  \left(16   f_1^2-16 f_2-h_1^2\right)}{32 } \nonumber\\
&&\pm\frac{3\sqrt{3} S_{2,1}^2}{8 \sqrt{b}},\\
S_{2,0}&=&   \pm\frac{1}{2}\sqrt{\frac{b}{3}}.\label{sols22}
\end{eqnarray}
Note that these expressions do not fix all the $S_{i,j}$. Still, this is enough to show that the relevant solutions exist. What is more, it can be explicitly verified that, for all the nontrivial solutions, $s^2=b+O(1/r)$, i.e., that these solutions tend to the bottom of the potential as $r\to \infty$.

These solutions must now be inserted in the modified Einstein equation \eqref{modEinstein_s}. The trivial solution, when inserted into the $tt$ component of Eq.~\eqref{modEinstein_s}, produces conditions that cannot be met under the working assumptions. The solutions \eqref{sols11}-\eqref{sols12} produce
\begin{equation}
0=    \frac{-24\pi b\mp\sqrt{\frac{3b}{2}}}{r^2}+O\left(\frac{1}{r^3}\right),
\end{equation}
which are also impossible to solve. Finally, the solutions \eqref{sols21}-\eqref{sols22} solve all the components of Eq.~\eqref{modEinstein_s} to $O(1/r^3)$ (despite the fact that not all the $S_{i,j}$ have been specified); therefore, there are static and spherically symmetric solutions that are also asymptotically flat. Moreover, the asymptotically flat solutions tend to the bottom of the potential as $r\to \infty$.

\section{Conclusions}\label{Conc}

\subsection{Interpretation}

In the previous sections it was found that there are vacuum solutions for $u$ and $s^{ab}$ that are static, spherically symmetric, and asymptotically flat, while these kind of solutions do not exist for $t^{abcd}$. The main difference in these cases is the relation between the different parts of the curvature tensor, namely, $R$, $R^{\rm T}_{ab}$ and $W_{abcd}$, and the coefficients, stemming from the coefficients' equations of motion. Notice that these relations are valid for any solution regardless of the symmetries it may have. Now, in the $u$ and $s^{ab}$ cases, the equation of motion, Eqs. \eqref{modEinstein_u}-\eqref{EOM_u} or \eqref{modEinstein_s}-\eqref{EOM_s}, do not depend on $W_{abcd}$. Therefore, $W_{abcd}$ can accommodate itself to have the appropriate asymptotic behavior.\footnote{$W_{abcd}$ has a very particular structure when approaching null infinity \cite{Penrose,Geroch}, reflecting the property that Weyl is conformally invariant.} In contrast, for $t^{abcd}$, there are fall off requirements on Ricci from the modified Einstein equation \eqref{modEinstein} and additional conditions on Weyl from the $t^{abcd}$ equation of motion, Eq.~\eqref{EOMt}.

In the particular case where the solutions are static and spherically symmetric, one may say that, for the $u$ and $s^{ab}$ coefficients, spacetime near $r\to \infty$ becomes Schwarzschild-like in the sense that the Ricci tensor is restricted to go to zero, while the Weyl tensor is not subject to a direct restriction. On the other hand, in the case of $t^{abcd}$ when subject to the same symmetries, all the pieces of the Riemann tensor have to go to zero, thus producing a Minkowski-like region near spatial infinity.

The question that arises is whether the impediments to find vacuum and asymptotically flat solutions in the $t^{abcd}$ case, under the assumptions that all the fields are static and spherically symmetric, are due to the potential or to the nonminimal coupling of $t^{abcd}$ and the Weyl tensor. To find clues in this direction one can study different potentials. It is particularly enlightening to analyze the (convex) potential obtained by setting $b=0$. It is clear from Eqs.~\eqref{modEinstein_s} and \eqref{EOM_s}, with $b=0$, that $S_1(r)\equiv 0\equiv S_2(r) $, which in turn implies that all the components of the Einstein tensor and the traceless Ricci tensor vanish. Thus, the Ricci tensor vanishes and, using Birkhoff's well-known theorem, it is possible to conclude that the solution in this case is a Schwarzschild spacetime. Something completely analogous happens in the $u$ case. However, the $t^{abcd}$ case is completely different: if one demands the solution to lie at the bottom of the potential, then $T(r)\equiv 0$. Therefore, the equations one obtains by writing Eqs.~\eqref{modEinstein} and \eqref{EOMt} with $b=0$ and using the metric and $t^{abcd}$ expressions that are compatible with the symmetries under consideration, imply that all the components of both, the Einstein tensor and the Weyl tensor, vanish. Thus, the solution is necessarily a flat spacetime.

This simple analysis reveals that, under the considered symmetries, spacetime can only be nontrivial if $t^{abcd}\neq 0$, and this is ultimately a consequence of the nonminimal coupling of $t^{abcd}$ and the Weyl tensor. Moreover, in a theory with $t^{abcd}$ but without the nonminimal coupling, the Weyl tensor does not appear in the equations of motion, avoiding all these problems. Consequently, the fact that the nonminimal coupling $t^{abcd}W_{abcd}$ in the action introduces a Weyl tensor in the equations of motion seems to be the reason for the obstructions appearing near spatial infinity that is found under the simplifying assumptions that the solutions are static and spherically symmetric.

Incidentally, the fact that static and spherically-symmetric solutions with $t^{abcd}=0$ require spacetime to be flat is closely related to the result shown in the Appendix \ref{blackholes} that, for a convex potential, every regular black hole solution that is static, spherically symmetric, and asymptotically flat, must be surrounded by a nontrivial configuration of SME coefficients, or, in the black hole argot, that every such black hole is hairy.

\subsection{Final thoughts}

The minimal gravity sector of the SME was studied under the assumption that all fields are static and spherically symmetric, in the case where Lorentz violation arises spontaneously. The goal was to study if it is possible to have solutions that are asymptotically flat. The result is that such solutions can exist for the $u$ and $s^{ab}$ coefficients, but no solutions of this form exist for the $t^{abcd}$ case.

This is probably one of the biggest clues to date on the $t$ puzzle: the fact that the $t^{abcd}$ term is the only part of the minimal gravity sector of the SME that produces no effects when a post-Newtonian approximation is used \cite{BaileyKostelecky2006}. In fact, in this post-Newtonian approximation, the asymptotically flat region $r\to \infty$ plays a key role as it is where the coordinates are defined. If the result found here is eventually shown to be generic, i.e., valid for nonsymmetric solutions, then the source of the $t$ puzzle would be revealed, ending many years of speculations. This could be done using the covariant definition of asymptotic flatness. However, this line of research seems daunting. Still, it would be relevant to verify if the results presented here hold in less symmetric situations.

Note that adding the $t^{abcd}W_{abcd}$ action term cannot be thought of as a small modification to general relativity, even if the components of $t^{abcd}$ have small values. This is because such a term implements additional restrictions on the Weyl tensor, which is the only piece of the curvature tensor that is not directly determined by the Einstein equation. This has profound consequences, as the analysis presented here shows. In addition, Eq.~\eqref{EOMt} suggests that, when the $t^{abcd}$ term is considered, there is no limit in which general relativity is recovered. Perhaps this is enough to disregard the $t^{abcd}$ term and other terms outside the minimal sector that may give rise to similar issues.

\begin{acknowledgments}
We acknowledge getting valuable feedback from Alejandro Aguilar, Marcelo Salgado, and Mike Seifert. This research was funded by UNAM-DGAPA-PAPIIT Grant IG-100120 and CONACYT (FORDECYT-PRONACES) Grant 140630.
\end{acknowledgments}

\bibliography{References}

\appendix
\section{Hamiltonian and stability of the theory}\label{Hamilto}

In this Appendix the Hamiltonian of the SME coefficients is studied to analyze when it is bounded from below, avoiding instabilities. For simplicity, this section only considers the Lagrangian \eqref{Lagk} in a flat spacetime and using Cartesian coordinates $(t,x,y,z)$. No coupling to gravity is considered. To reflect the fact that the calculations are done in a particular coordinate system, the spacetime indexes are Greek. In addition, Latin indexes from the mid alphabet represent spatial indexes associated with $x,y,z$ and the metric components are $\eta_{\mu\nu}={\rm diag}(-1,1,1,1)$. Again, the metric and its inverse $\eta^{\mu\nu}={\rm diag}(-1,1,1,1)$ are used to lower and raise Greek indexes.

Under this circumstances the Lagrangian \eqref{Lagk} can be written as
\begin{equation}
    \mathcal{L}_k(k)=\frac{1}{2}\partial_t k_{\mu\nu\rho\sigma}\partial_t k^{\mu\nu\rho\sigma}-\frac{1}{2}\partial_i k_{\mu\nu\rho\sigma}\partial^i k^{\mu\nu\rho\sigma}-V(k^2).
\end{equation}
Clearly the conjugate momenta is
\begin{equation}
   \pi_{\mu\nu\rho\sigma}=\partial_t k_{\mu\nu\rho\sigma}.
\end{equation}
Thus, the Hamiltonian takes the form
\begin{eqnarray}
    \mathcal{H}_k(k)&=&\pi_{\mu\nu\rho\sigma}\partial_t k^{\mu\nu\rho\sigma}-\mathcal{L}_k(k)\nonumber\\
    &=&\frac{1}{2}\pi_{\mu\nu\rho\sigma}\pi^{\mu\nu\rho\sigma}+\frac{1}{2}\partial_i k_{\mu\nu\rho\sigma}\partial^i k^{\mu\nu\rho\sigma}+V(k^2).\nonumber\\
    &&\label{Hamiltonian}
\end{eqnarray}

The potentials under consideration are non-negative, therefore, the last term of the Hamiltonian cannot produce instabilities. Unfortunately this is not the case for the first two terms in $ \mathcal{H}_k(k)$. In fact, these terms have no definite sign as, when expanded to perform the index addition, they produce nonpositive terms. The nonpositive terms have an odd number of $t$ indexes (which have and odd number of $\eta_{tt}=-1$). However, under the assumption that the fields are static, these nonpositive terms in the Hamiltonian vanish. Hence, the Hamiltonian given in Eq.~\eqref{Hamiltonian} is bounded from below provided that the solutions are strictly static, which is the case studied in this paper.

The fact that generic perturbations can produce instabilities is clearly unsatisfactory. Moreover, it seems that the nonminimal coupling term in the action \eqref{action} can also produce instabilities. Unfortunately a rigorous study of the theory's stability lies outside the present scope of our work and it is left for the future.

\section{Black hole solutions with convex potentials}\label{blackholes}

No-hair theorems are rigorous results showing that there are no nontrivial configurations for some matter fields around a black hole. A striking feature of the no-hair theorems is that, generically, to prove them there is no need to solve the equations of motion. Perhaps the simplest no-hair proof, which is valid for a real scalar field whose kinetic term is a Klein-Gordon term and it is subject to a convex potential, was proposed by Bekenstein \cite{PhysRevLett.28.452,PhysRevD.5.2403}, and this method is generalized here. Other well-known no-hair theorems can be found in Refs. \onlinecite{PhysRev.164.1776,Israel1968,PhysRevLett.26.331,PhysRevLett.26.1653,Sudarsky_1995,PhysRevLett.99.201101} (see Refs.~\onlinecite{Chruciel2012,Cardoso2016} for extensive reviews).

In this Appendix, the theory under consideration is given by the action \eqref{action}, namely, all the pieces of $k^{abcd}$ are considered. However, the potential $V=V(k^2)$ is kept arbitrary for the moment. In vacuum, the equations of motion are
\begin{eqnarray}
G_{ab}&-&\frac{g_{ab}}{2}k^{cdef}R_{cdef} - R_{cde(a}{k^{cde}}_{b)}\nonumber\\
&&- 2 \nabla_{c} \nabla_{d} {{k^c}_{(ab)}}^d=8\pi T_{ab}, \label{modEinstein_k}\\
R_{abcd}&=&16\pi \left(-\nabla_e \nabla^e k_{abcd} + 2 V' k_{abcd}\right).\label{EOM_kcompleto}
\end{eqnarray}
where
\begin{eqnarray}
T_{ab}
&=&g_{ab}\left[-\frac{1}{2}(\nabla k)^2-V\right] + \nabla_a k^{cdef} \nabla_b k_{cdef}\nonumber\\
&&+4\nabla_c\left[ {k_{def}}^c \nabla_{(a}{k^{def}}_{b)}- k_{def(a}\nabla_{b)}k^{defc}\right]\nonumber\\
&&+4 k_{def(a}\nabla_c\nabla^c {k^{def}}_{b)}-8V' k_{cde(a}{k^{cde}}_{b)},\nonumber\\
&& \label{Tab_k}
\end{eqnarray}
and $V'$ is the potential derivative with respect to $k^2$.
The idea is to study static, spherically symmetric black hole solutions to these equations of motion that are also regular and asymptotically flat.

Following Bekenstein, the first step is to contract the equation of motion for $k^{abcd}$, Eq.~\eqref{EOM_kcompleto}, with $k^{abcd}$, which leads to
\begin{eqnarray}
R_{abcd}k^{abcd}&=&16\pi \left(-k^{abcd}\nabla_e \nabla^e k_{abcd} + 2 V' k^2\right)\nonumber\\
&=&16\pi \left[-\nabla_e (k^{abcd}\nabla^e k_{abcd})+(\nabla k)^2 + 2 V' k^2\right],\nonumber\\
&&\label{kR1}
\end{eqnarray}
where a divergence is constructed. 

Another expression for $R_{abcd}k^{abcd}$ can be obtained from the trace of Eq.~\eqref{modEinstein_k}, which takes the form
\begin{equation}
-R-3k^{abcd}R_{abcd} - 2 \nabla_{b} \nabla_{c} {{k^b}_{a}}^{ac}=8\pi T, \label{kR2}
\end{equation}
where
\begin{eqnarray}
T&\equiv&g^{ab}T_{ab}\nonumber\\
&=&-5(\nabla k)^2-4V -8V' k^2+\nabla_c\left[ 2\nabla^c k^2\right.\nonumber\\
&&+ \left.4{k_{def}}^c \nabla_a k^{defa}  - 4k^{defa}\nabla_a {k_{def}}^c\right].
\label{traceT}
\end{eqnarray}
Equating the two expressions for $k^{abcd}R_{abcd}$ produces
\begin{eqnarray}\label{EqMas}
0&=&\frac{R}{8\pi}+ \nabla_a \alpha^a +(\nabla k)^2 + 4 V' k^2-4V,
\end{eqnarray}
with
\begin{eqnarray}
\alpha^a &\equiv&\frac{1}{4\pi} \nabla_{c} {{k^a}_{b}}^{bc} -6 k^{bcde}\nabla^a k_{bcde}+ 2\nabla^a k^2 \nonumber\\
&& +4{k_{cde}}^a \nabla_b k^{cdeb} - 4k^{cdeb}\nabla_b {k_{cde}}^a. \label{alpha}
\end{eqnarray}

Equation \eqref{EqMas} is the key for this study, which now calls for the implementation of the symmetries. For $k^{abcd}$ to be static and spherically symmetric it requires that all its irreducible pieces are compatible with these symmetries. Under these assumptions
\begin{equation}
k^2=12 (f hT)^2+\frac{1}{2} (fS_1)^2+\frac{1}{2} (h S_2)^2+\frac{1}{4}(f S_1- h S_2)^2+ 6 u^2,\label{ksqared}
\end{equation}
and
\begin{eqnarray}
 (\nabla k)^2 &=& \frac{72f^2hT^2}{r^2}   + \frac{12}{h}  \left[ \frac{d(fhT)}{dr} \right]^2 \nonumber\\
 &&+ \frac{( f S_1-3hS_2 )^2}{2 r^2 h}  +   \frac{(fS_1+h S_2)^2 }{4 f^2 h}\left( \frac{df}{dr}\right)^2 \nonumber \\
  & & + \frac{1}{4 h} \left[ \frac{d ( fS_1-h S_2)}{dr}\right]^2 + \frac{1}{2h}\left[ \frac{d( f S_1)}{dr}  \right]^2       \nonumber \\
 & & + \frac{1}{2h} \left[ \frac{d(hS_2)}{dr}   \right]^2 + \frac{6}{h} \left(\frac{du}{dr}\right)^2.\label{Dk2}
\end{eqnarray}
Given that $r>0$, and $f>0$ and $h>0$ (as attention is restricted to the black hole exterior), it becomes clear that $k^2\geq 0$ and $ (\nabla k)^2\geq 0$.

The integral of Eq.~\eqref{EqMas} in a spacetime region $\mathcal{B}$ takes the form
\begin{eqnarray}
0&=&\int_\mathcal{B} d^4 x \sqrt{-g} \left[ \frac{R}{8\pi}+(\nabla k)^2 + 4(V'   k^2 -V)+\nabla_a \alpha^a\right]\nonumber\\
&=&\int_\mathcal{B} d^4 x \sqrt{-g} \left[ \frac{R}{8\pi}+(\nabla k)^2 + 4(V'   k^2 -V)\right]\nonumber\\
&&+\oint_\mathcal{\partial \mathcal{B}} d^3 x \sqrt{\gamma} n_a  \alpha^a,\label{integralBek}
\end{eqnarray}
where in the last line the pseudo-Riemannian version of Stokes theorem is used \cite[Appendix B]{Wald} and $n^a$ is the unit vector normal to the boundary of $\mathcal{B}$, which is denoted by $\partial \mathcal{B}$.

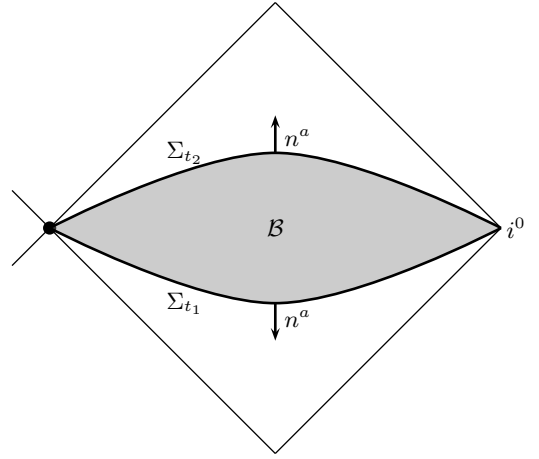
\begin{figure}
\centering
\begin{pspicture}(6,6.5)
\psline[linewidth=.5pt](-0.5,2.5)(3,6)
\psline[linewidth=.5pt](3,0)(-0.5,3.5)
\psline[linewidth=.5pt](6,3)(3,6)
\psline[linewidth=.5pt](3,0)(6,3)
\psdot[dotsize=5pt](0,3)
\pscustom[linewidth=1pt,fillcolor=black!50,fillstyle=solid,opacity=0.4]{%
\pscurve[linewidth=1pt](0,3)(3,2)(6,3)
 }
\pscustom[linewidth=1pt,fillcolor=black!50,fillstyle=solid,opacity=0.4]{%
\pscurve[linewidth=1pt](0,3)(3,4)(6,3)
 }
\psline[arrows=->,linewidth=1pt](3,4)(3,4.5)
\psline[arrows=->,linewidth=1pt](3,2)(3,1.5)
\rput(3,3){$\mathcal{B}$}
\rput(3.3,4.2){$n^a$}
\rput(3.3,1.8){$n^a$}
\rput(1.8,4){$\Sigma_{t_2}$}
\rput(1.8,2){$\Sigma_{t_1}$}
\rput(6.2,3){$i^0$}
\end{pspicture}
\caption{Penrose diagram for the exterior of the black hole where the spacetime region $\mathcal{B}$ is marked on gray. The boundary of $\mathcal{B}$ consists of four pieces: two constant $t$ hypersurfaces, $\Sigma_{t_1}$ and $\Sigma_{t_2}$, the bifurcation surface, represented by a black dot, and spatial infinity, $i^0$. The unit normal vectors for $\Sigma_{t_1}$ and $\Sigma_{t_2}$, $n^a$, are also represented.}
\label{PenroseDiag}
\end{figure}

Provided that spacetime is a regular black hole, the horizon contains a bifurcation surface where, loosely speaking, all constant $t$ hypersurfaces $\Sigma_ t$, intersect. Recall that $t$ is the coordinate along the timelike Killing vector field. Taking these conditions into the account, it is possible to verify that, to have a vanishing boundary contribution in Eq.~\eqref{integralBek}, $\mathcal{B}$ can be chosen as the region between any $\Sigma_{t_1}$ and $\Sigma_{t_2}$ (with $t_1<t_2$). In Fig.~\ref{PenroseDiag} a Penrose diagram of the black hole exterior is presented and the region $\mathcal{B}$ is illustrated. The boundary of $\mathcal{B}$ contains four pieces: the bifurcation surface, $\Sigma_{t_1}$, $\Sigma_{t_2}$, and $i^0$. Now, the contribution of the bifurcation surface vanishes because this surface is of zero measure (it is two dimensional) and the contribution at $i^0$ vanishes because the integrand goes to zero by virtue of asymptotic flatness. The contributions of $\Sigma_{t_1}$ and $\Sigma_{t_2}$, which do not vanish, are of the same size but have the opposite signs: the size is the same since the solution is static and the sign difference stems from the fact that, according to Stokes theorem, $n^a$ points outwards from $\mathcal{B}$ (see Fig.~\ref{PenroseDiag}). Therefore, these two integrals cancel, the boundary integral vanishes for any region $\mathcal{B}$ constructed in this way, and Eq.~\eqref{integralBek} becomes
\begin{equation}
0=\int_\mathcal{B} d^4 x \sqrt{-g} \left[ \frac{R}{8\pi}+(\nabla k)^2 + 4(V'   k^2 -V)\right].\label{intNoBound}
\end{equation}

The next goal is to find the conditions for the third term in the integrand of Eq.~\eqref{intNoBound} to be non-negative. Assuming
\begin{equation}
V=\sum_{N=0}^\infty c_N (k^2)^N,
\end{equation}
where $c_N$ are constants, then
\begin{equation}
V' k^2-V=\sum_{N=0}^\infty (N-1) c_N (k^2)^N.
\end{equation}
Therefore, for this term to be non-negative one needs $c_0\leq 0$ and $c_N\geq 0$ for $N=2,3,\ldots$. Note that $c_1$ can be arbitrary, and that, $c_0$ acts as a cosmological constant.\footnote{The potential \eqref{potential} used for spontaneous Lorentz violation has positive $c_0$, and thus, it is not of the type considered here.}

Suppose now that the potential is such that $V' k^2-V\geq 0$ and $c_0=0$ so that spacetime is asymptotically flat. Then, one can demonstrate that either the exterior region is flat, which does not correspond to a conventional black hole, or there exist at least one open $r$ interval where $R<0$. To see this assume $R\geq0$ for all $r$ in this region. Then the integral~\eqref{intNoBound} is a sum of three non-negative terms that add up to zero. Therefore each of these terms must vanish, including $R$. It then follows that $(\nabla k^2)=0=k^2$, and from Eq.~\eqref{EOMt} it becomes clear that $R_{abcd}=0$, namely, the exterior is flat. The other possibility is that $R$ is such that its $r$ integral (with the corresponding volume element) over the exterior region is negative. In this last case, there has to be a nonvanishing $k^{abcd}$ configuration for the integral~\eqref{intNoBound} to vanish. Therefore, in the theory under consideration, every regular and asymptotically flat black hole subject to the symmetries that are assumed throughout the text, needs to have a region where $k^{abcd}\neq 0$. Interestingly, this could allow gravity tests that use black hole shadows \cite{PhysRevLett.125.141104} to place limits on theories with nonminimal couplings of curvature and (dynamical) tensor fields.

\end{document}